# Quantification of internal dosimetry in PET patients II: Individualized Monte Carlo-based dosimetry for [18F]Fluorocholine PET

Sara Neira[1], Jacobo Guiu-Souto[2], Paulino Pais[3], Sofía Rodríguez Martínez de Llano[3], Carlos Fernández[2], Virginia Pubul[4], Álvaro Ruibal[4,5,6,7], Miguel Pombar[5,8], Araceli Gago-Arias[1,8,9], Juan Pardo-Montero[1,8,*]

1. Group of Medical Physics and Biomathematics, Instituto de Investigación Sanitaria de Santiago, Travesía Choupana s/n, 15706 Santiago de Compostela, Spain.

2. Department of Medical Physics, Centro Oncolóxico de Galicia, C/ Doctor Camilo Veiras 1, 15009 A Coruña, Spain.

3. Department of Nuclear Medicine, Centro Oncolóxico de Galicia, C/ Doctor Camilo Veiras 1, 15009 A Coruña, Spain.

4. Department of Nuclear Medicine, Complexo Hospitalario Universitario de Santiago de Compostela, Travesía Choupana s/n, 15706 Santiago de Compostela, Spain.

5. Group of Molecular Imaging and Oncology, Instituto de Investigación Sanitaria de Santiago, Travesía Choupana s/n, 15706 Santiago de Compostela, Spain.

6. Molecular Imaging Group, Department of Radiology, Faculty of Medicine, Universidade de Santiago de Compostela, Campus Vida, 15782 Santiago de Compostela, Spain.

7. Fundación Tejerina, C/ José Abascal 40, 28003 Madrid, Spain.

8. Department of Medical Physics, Complexo Hospitalario Universitario de Santiago de Compostela, Travesía da Choupana s/n, 15706 Santiago de Compostela, Spain.

9. Institute of Physics, Pontificia Universidad Católica de Chile, Santiago, Chile.

* **Corresponding author:** Juan Pardo-Montero, Instituto de Investigación Sanitaria de Santiago (IDIS), Servizo de Radiofísica e Protección Radiolóxica, Hospital Clínico Universitario de Santiago, Trav. Choupana s/n, 15706, Santiago de Compostela; Phone: +34 981955604; E-mail: juan.pardo.montero@sergas.es

**Running title:** Monte Carlo dosimetry of FCH-PET

**Conflict of interest:** The authors declare that they have no competing interests.

# Abstract

**Purpose:** To obtain individualized internal doses with a Monte Carlo method in patients undergoing diagnostic [18F]FCH-PET studies and to compare such doses with the MIRD method calculations.

**Methods:** A patient cohort of 17 males were imaged after intravenous administration of a mean [18F]FCH activity of 244.3 MBq. The resulting PET/CT images were processed in order to generate individualized input source and geometry files for dose computation with the MC tool GATE. The resulting dose estimates were studied and compared to the MIRD method with two different computational phantoms. Mass correction of the S-factors was applied when possible. Potential sources of uncertainty were closely examined: the effect of partial body images, urinary bladder emptying, and biokinetic modeling.

**Results:** Large differences in doses between our methodology and the MIRD method were found, generally in the range of ±25%, and up to ±120% for some cases. The mass scaling showed improvements, especially for non-walled and high-uptake tissues. Simulations of the urinary bladder emptying showed negligible effects on doses to other organs, with the exception of the prostate. Dosimetry based on partial PET/CT images (excluding the legs) resulted in an over-estimation of mean doses to bone, skin, and remaining tissues, and minor differences in other organs/tissues. Estimated uncertainties associated with the biokinetics of FCH introduce variations of cumulated activities in the range of ±10% in the high-uptake organs.

**Conclusions:** The MC methodology allows for a higher degree of dosimetry individualization than the MIRD methodology, which in some cases leads to important differences in dose values. Dosimetry of FCH-PET based on a single partial PET study seems viable due to the particular biokinetics of FCH, even though some correction factors may need to be applied to estimate mean skin/bone doses.

# 1. Introduction

Quantitative imaging PET/CT by radiolabeled [18F]Fluorocholine (FCH) has become an important tool for prostate cancer diagnosis[1]. It is especially useful for re-staging patients with biochemical relapse after radiotherapy treatment. It has been shown that PET/CT with [18F]FCH is more sensitive than [18F]Fluoro-2-deoxy-2-d-glucose (FDG) to detect metastatic lymph nodes or bone metastatic lesions[2]. Its use for prostate cancer diagnosis is based on the high metabolic rate of tumor cells that require choline for the biosynthesis of phospholipids[3]. FCH also has other applications, for example for the detection of brain cancer, where FDG is not recommended due to the high uptake of the brain[4].

Internal dosimetry in medical procedures with radiopharmaceuticals is shifting from populational-based to individualized methods, pushed by new legislation in some countries, like the European Council directive 2013/59/Euratom in its Article no.56[5]. Dosimetric calculations are in practice usually based on average population tables. For instance, in diagnostic applications of [18F]FCH, the ICRP 128's[6] recommends dosimetric estimations based on tabulated coefficients of absorbed doses per injected activity. Moreover, it does not provide solutions for individualization beyond a classification depending on age. These tables are obtained from the application of the fast, simple, and well-established Medical Internal Radiation Dose (MIRD) formalism[7,8]. The set of dose-activity coefficients is pre-calculated by Monte Carlo (MC) simulation of population-averaged geometries and homogeneous-in-organ activity distributions. Dose estimates can be improved if patient-specific activity distributions are employed, allowing for a higher degree of individualization than the ICRP coefficients alone. Yet, the geometries remain generic, and although numerous families of computational phantoms have been developed over the years[9], differences with actual patients' bodies will inevitably be present.

It has already been shown that considerable uncertainties can arise from the use of generic activity distributions and non-personalized geometries – The MIRD pamphlet no. 11[10] remarked the variations of the Specific Absorbed Coefficients (SAFs) due to mass discrepancies between real patients and computational phantoms as a limitation. There, a SAF correction method for photons and beta particles was provided, which is incorporated in the well-known OLINDA/EXM software[11]. Several studies have further validated the mass correction formulas, while also characterized the differences that might arise from employing non-individualized geometries. Petoussi-Henss et al.[12] obtained results that were compatible with the mass-scaling formulas for most source-target pairs, but they also highlighted the interorgan distance as another source of discrepancies for the cross-irradiation photonic SAFs. Two complementary studies by Marine et al.[13] and Clark et al.[14] quantified the uncertainty of the photonic SAFs in a family of computational phantoms due to geometric variations of different body types and obesity levels. The loss of the heterogeneity of activity distributions constitutes another drawback of the MIRD formalism. Several studies have reported considerable differences between a direct Monte Carlo dosimetry approach versus the use of computational phantoms[15,16]. A recent publication by our group for FDG-PET was consistent with these results[17].

The use of non-individualized dosimetric methods has an impact both on diagnostic and therapeutic applications with radiopharmaceuticals. In the case of diagnostic imaging, although the levels of radiation damage received by the patient in one PET study are relatively low, accurate dosimetry is interesting from the point of view of radiation protection (especially for those patients that recurrently undergo studies as part of some prolonged treatment or monitoring). In the case of therapeutic applications, accurate dosimetry methods are important for the development of treatment planning systems, which would allow for personalized treatment optimization. In addition, the consideration of uptake heterogeneities would allow to obtain realistic dose-volume histograms (DVH), which are paramount for the evaluation of dose distributions, especially in tumors[18].

The main aim of this work is to provide a fully-personalized PET dosimetric estimate for patients undergoing diagnostic PET/CT studies with [18F]FCH. The differences between this approach and both the MIRD method and the current ICRP recommendations are characterized, including patient-specific information in the MIRD method whenever possible. We also evaluate different sources of uncertainty and study the limitations of performing dosimetry calculation from common PET/CT protocols, which typically include static studies and partial images. This piece of work can be seen as a continuation of our previous study[17], where we presented an individualized, MC-based dose computation methodology, and applied it to a cohort of FDG patients. We shall mention other works that have presented similar platforms/methodologies for individualized MC dose computation for radionuclides, as RAPID[19], VIDA[20], OEDIPE[21], or RAYDOSE[22].

## 2. Materials and Methods

### 2.1. Patient Cohort

We recruited 17 male patients, with a mean age of 67 years (range 54-83 years) and an average body mass index (BMI) of 29.1 kg/m$^2$ (range 22.6-37.9 kg/m$^2$). They were prostate cancer patients, of which 65% (11) had undergone radical prostatectomy. Mean activity of 244.3 MBq (range 166.1-297.9 MBq) of [18F]FCH was administered intravenously.

### 2.2. PET/CT equipment and imaging protocol

PET/CT images were obtained with a hybrid system (Ingenuity TF PET/CT, Philips Healthcare, Best, Netherlands). This PET unit implements a Time-of-Flight technology with an axial Field-of-View of 18 cm and uses 4×4×22 mm$^3$ LYSO crystal detectors. The CT presents a helical acquisition mode with 4 cm of axial coverage and 64-detectors.

The acquisition protocol comprised a dual-phase procedure. The first stage consisted of an examination of the pelvic region (1 bed position) in order to evaluate the interference of radioactive urine in the excretory pathways. A barium oral contrast was administered to patients 15 minutes before the CT. The PET acquisition started immediately after the [18F]FCH injection. However, this first PET/CT (which is part of the clinical protocol) was not employed during this study, because being acquired this early, no urinary bladder contents could be measured, nor other tissue like the kidneys, which were only partially imaged. In the second stage, performed on average 70 min later (range 30-100 min), patients were invited to empty their bladder, and afterward an image was acquired from the top of the skull to the upper part of the legs. PET images were acquired in 3D mode and reconstructed by the ordered subset expectation maximization likelihood algorithm of the manufacturer after attenuation correction based on the CT. Time per bed position was set to 3 min and image resolution was set to 144×144 pixels, with a slice thickness of 4 mm. In both phases, patients were placed in the supine position, with the arms crossed above the head. The CT protocol consisted of a helical scan with 120 kV, a slice thickness of 3 mm, and an acquisition matrix of 512×512 pixels. The parameters of the reconstruction of the images were set to 3 iterations, 33 subsets, a kernel width of 14.1 mm, and a relaxation parameter value of 1. No PSF correction was applied.

**2.3. Segmentation**

Several organs and tissues of interest were manually segmented on the second CT image of each patient (Supplementary Materials, Table SM1, for a complete list of organs/tissues and abbreviations). We used the software 3DSlicer[23] (www.slicer.org), which includes several useful semi-automatic tools to assist with the task. The urinary bladder (UB) was semi-automatically segmented with the Chan-Vese method[24]. The initial contour input was manually drawn over the CT image, while the active contours algorithm was run over the PET image in Matlab (The MathWorks, Natick, MA). This procedure was similar to that used in our previous work[17].

### 2.4. Reconstruction of cumulated activities

#### 2.4.1. Biokinetic Modeling

A biokinetic model has to be used to obtain individualized time-activity curves (TACs) for each patient. Currently, for [18F]FCH the ICRP 128 proposes a tabulated biokinetic model that includes the main tissues involved in the biokinetics of FCH[6]. However, it is a population-averaged model that allows for little customization on a patient-specific basis. The ICRP 128 proposal is based on data from several studies[25–27], where the biodistribution of FCH was thoroughly measured for different patient cohorts. Of special interest is the study by Giussani et al.[25], where a compartmental biokinetic model was developed and fitted to experimental activities. However, the measurement of activity at several time-points in a realistic scenario is rarely performed. The fit of one time-point per compartment to the model would likely lead to the degeneration of the predicted TACs.

Several studies have suggested that the TACs of FCH might be well approximated by a simple exponential model[27,28]. We have followed this approximation to construct TACs in this work, although some calculations were performed in order to quantify the deviations that might be caused by this approximation compared to more complex models presented within the framework of the MADEIRA project[25,29]. Therefore, cumulated activities (CA) are calculated by correcting by decay the activity in the PET image to the time of the injection ($t=0$) and integrating the initial activity distribution in the range $[0, \infty)$:

$$CA_i = \int A_i^0 e^{-\lambda t} dt = \frac{A_i^0}{\lambda} = \frac{A_i^{PET} e^{\lambda t_{PET}}}{\lambda} \qquad (1)$$

where $i$ is the i-th voxel of the image, $\lambda=1.05\times10^{-4}$ s$^{-1}$ is the 18-F decay constant, $t_{PET}$ is the time-point to which the PET image was corrected (start of acquisition) with respect to the time of the injection, and $A_i^{PET}$ is the activity distribution measured with the PET. This approach for modeling the pharmacokinetics of the radioisotope assumes that the initial uptake in each tissue is instantaneous and that there is no biological exchange nor clearance over time.

In order to justify the election of a simple exponential model and to investigate possible systematic deviations in the computation of cumulated activities, we characterized the differences between Giussani's biokinetic model, a biexponential fit, and the simple exponential model implemented in this study. With this aim, the experimental data reported in Giussani's work[25] were imported (data for 10 patients). We analyzed the liver, kidneys, and spleen.

### 2.4.2. Urinary Bladder emptying

Urinary bladder activities measured in the PET is excreted at some point. As urinary excretion was not measured for each patient, the dose in that organ and nearby organs may not be accurately estimated. In order to evaluate the effect of bladder emptying on the mean dose estimate, the CA map was modified to include a scenario where the bladder is emptied 10 min after the PET study ends and does not refill again. The activity in those voxels that belong to the UB contents was computed in this case as:

$$CA_{i \in UB} = \int_0^{t_{empty}} A^0_{i \in UB} e^{-\lambda t} dt = \frac{A^{PET}_{i \in UB} e^{\lambda t_{PET}}}{\lambda} \left(1 - e^{-\lambda t_{empty}}\right) \quad (2)$$

where $t_{empty}$ is the time at which it is considered that the contents of the bladder were fully excreted. The two scenarios presented here correspond to a static bladder with a patient-variable activity content (that measured by the PET) which in one case is never excreted and in the other is eliminated "as soon as possible".

## 2.5. Whole-body PET/CT versus partial images

As mentioned in Section 2.2, only partial PET/CT images were available. In particular, the lower part of the legs and the forearms/hands are not imaged. Activity concentration in the legs/arms is expected to be low and not to contribute much to doses elsewhere, yet we have investigated the possible effect of such missing activities.

From the last slice of the thighs, new tissue was created by extruding and thinning the leg tissue (simulating the thickness variation from thigh to ankle), including RT, bone and skin. The length of the artificial legs was extended to match the height of the patient reported in the PET/CT metadata. The new HU and activity were extrapolated from the last original PET/CT slices. We run simulations with and without artificial legs and compared mean doses for each organ. This procedure is presented in more detail in the Supplementary Materials, including **Figure SM1,** where we present slices of the reconstructed legs for one patient.

To validate this procedure, we also applied it to two whole-body FDG-PET/CT patients studied in our previous work[17]: in those patients, legs were removed from the mid-thigh, artificial legs were created and activities were assigned. Mean doses were computed and compared for the three cases: with original or rebuilt legs, and without legs.

In order to account for the missing activity of the forearms/hands, we connected both arm sections by a semi-torus by a strategy similar to the leg case. This rough approximation to the real geometry allowed estimating the expected deviations of the doses. The full procedure is detailed in the Supplementary Materials.

**2.6. Monte Carlo Simulations**

The MC software vGATE v9.0[30] (www.opengatecollaboration.org) was used to compute internal dose distributions. GATE is employed in medical imaging and radiotherapy applications and has already has been widely validated for internal dosimetry calculations[31,32]. It constitutes a flexible tool with many useful applications for internal dosimetry calculations, such as DICOM compatibility. Additional tools for pre- and post-processing were implemented in MATLAB (The Mathworks, Natick, MA).

The first step consisted in calculating the individual geometry and activity source files for each patient, which were exported into DICOM format with CT and PET dimensions, respectively. The geometry consisted of the original CT image with a few modifications: due to the administration of the barium contrast, stomach and intestine tissues often presented high Hounsfield units (HU); in order to avoid the GATE HU-to-material translator to identify them as bone tissue, those voxels were corrected to approximately match the density of water. The translation of HU to densities was performed by using the tables provided by GATE, setting a density tolerance of 0.1 g/cm$^3$. The CT original slices were reduced by 50%, to a slice spacing of 3 mm to avoid excessive calculation time. The source was calculated as described in Section 2.4, where each voxel accounts for the number of disintegrations that occurred within it (cumulated activity). Two different sources were created with and without bladder emptying, as described in 2.4.2.

GATE files were configured with the physics package emstandard_opt4, and cutoffs of 10 keV for photons and electrons. The GATE 18F built-in source (ion particle) was used, and a fixed number of $10^8$ primaries (18F disintegrations) was simulated for each patient. This number of primaries was large enough to obtain mean dose (the metric under study in this work) statistical uncertainties below 1% for most organs, calculated from voxel uncertainties by error propagation. Simulations were run in a virtual machine of 20 cores and 64 GB of memory, hosted by CESGA (Galician Supercomputing Centre). Each full simulation lasted approximately 100 h/core, and they were split into several cores in order to speed up the calculation time. This parallel computation produced several dose files, that were merged in a single file by considering the propagation of statistical uncertainties associated with the dose computation[33]. As the number of simulated disintegrations did not match the total cumulated activity, the dose had to be linearly scaled by a factor $CA_{total}/N_{primaries}$. Mean doses were calculated for each organ by employing the organs segmentations.

**2.7. Computational Phantoms**

We compared our methodology with the MIRD method. Two computational phantoms were chosen: the Cristy-Eckerman adult man[34] (CE), and the recently proposed male phantom of the ICRPs 110 and 133[35,36].

Essentially, in the MIRD formalism, the cumulated activities in each organ/tissue (with no heterogeneities) are multiplied by dose conversion factors (S-factors) associated with a specific geometry (phantom). More details about the MIRD method can be found elsewhere[7,8]. To account for organ mass differences between real patients and phantoms, S-factors can be corrected by adding patient-specific information[11]. The self-irradiation mass-scaled specific absorbed fractions (SAFs) for beta and gamma are:

$$\Phi_\beta(r_T \leftarrow r_T)' = \Phi_\beta(r_T \leftarrow r_T) \frac{m_{r_T}^{phantom}}{m_{r_T}^{patient}} \quad (3)$$

$$\Phi_\gamma(r_T \leftarrow r_T)' = \Phi_\gamma(r_T \leftarrow r_T) \left( \frac{m_{r_T}^{phantom}}{m_{r_T}^{patient}} \right)^{2/3} \quad (4)$$

where $\Phi_{\beta,\gamma}$ represents the SAFs of beta and gamma radiation, and $m_{r_T}^{phantom/patient}$ is the mass of the organ of the phantom or patient.

# 3. Results

**3.1. Analysis of different sources of uncertainty**

**3.1.1. Biokinetic model uncertainty**

In **Figure 1** we present differences in TACs and CAs between the exponential, biexponential, and Giussani's biokinetic models, for liver, kidneys, and spleen (experimental data taken from[25]). Cumulated activities median differences between the exponential and biexponential fits reach values of 0.05%, -4.6%, and -4.7% for liver, kidneys, and spleen, respectively. Differences between the exponential and Giussani's biokinetic models are -8.1%, -0.72%, and 3.4% for the same tissues.

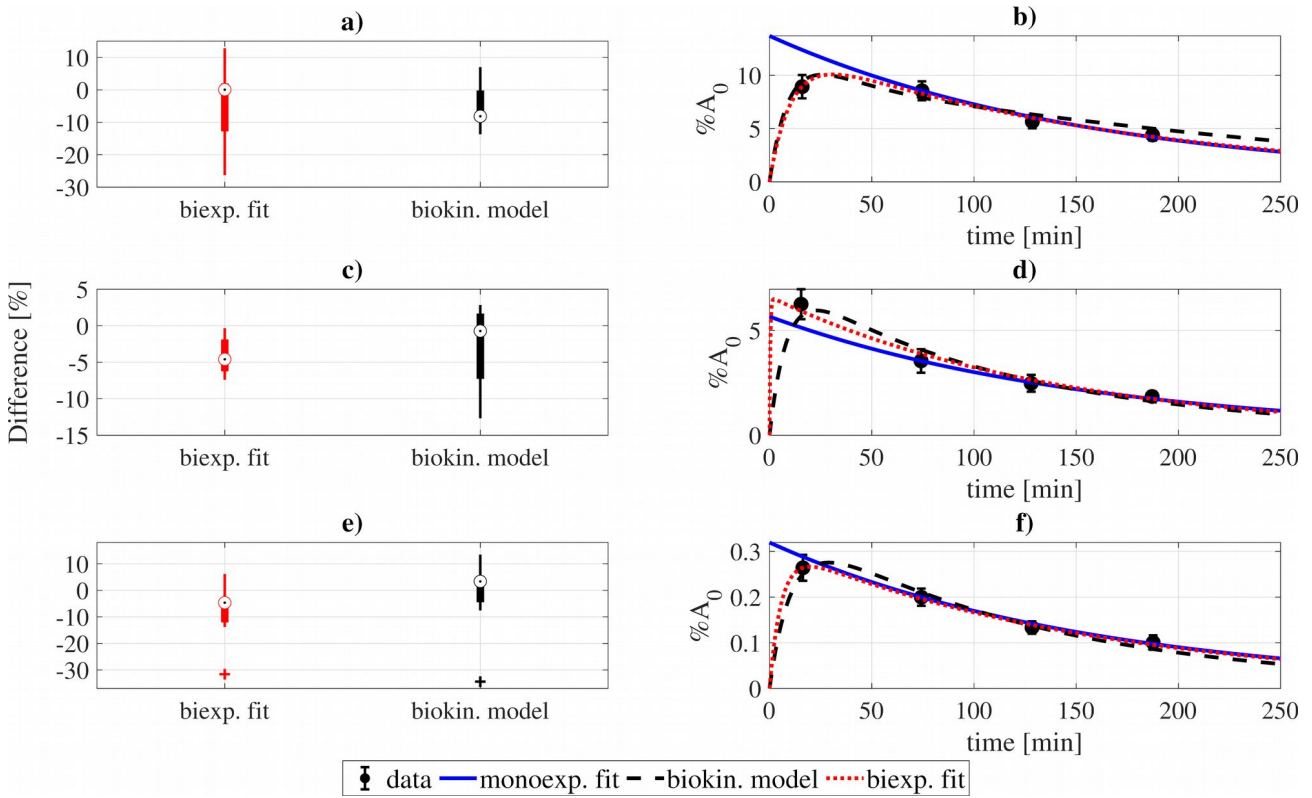

**Figure 1:** Evaluation of the cumulated activities (CA) obtained with three different fits to the experimental data for Guissani's patient cohort. Left panels: comparison of CA predicted by the monoexponential fit versus a biexponential fit and the biokinetic model ($CA_{exp}-CA_{model}$)/$CA_{model}$) for, a) liver, c) kidneys, and e) spleen. Right panels: Example of fits for b) liver, d) kidneys, and f) spleen.

### 3.1.2. Urinary bladder emptying

In **Figure 2A** we present the effect of bladder emptying at 10 min after the end of the PET scan or no emptying on mean doses in our patient cohort. The largest difference is for the UB wall, with a median value of $5.27\times10^{-3}$ mGy/MBq (range [$2.91\times10^{-3}$, $2.18\times10^{-2}$]), approximately a 30% relative difference between medians. The prostate also presents non-negligible differences between both cases, with median differences of $7.10\times10^{-4}$ mGy/MBq (range [$3.25\times10^{-4}$, $2.14\times10^{-3}$]), approximately 5% difference between medians. Doses to most organs are not affected by bladder emptying.

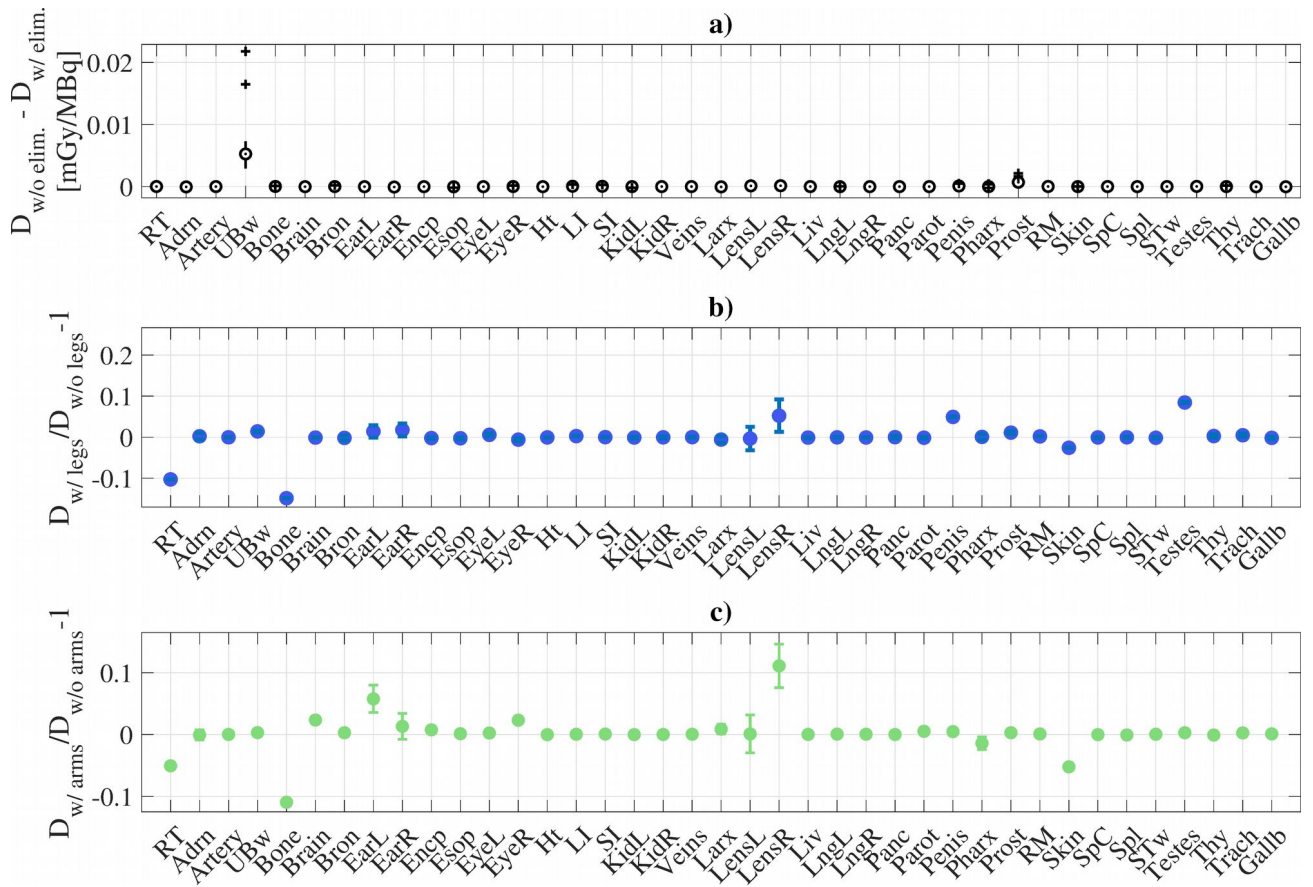

**Figure 2:** Variation of mean doses per organ due to different effects/limitations: a) Effect of urinary bladder emptying (with and without elimination) in terms of dose per unit of injected activity for the entire patient cohort; effect of legs (b) and arms (c) reconstruction in three patients, expressed as the mean ±1 standard deviation of relative differences.

### 3.1.3. The effect of partial PET/CT images

A simulation where legs were artificially reconstructed for the source and geometry was run for three patients. Organ mean doses were compared to the case of the originally incomplete images. Mean doses to the majority of organs are not affected by the addition of the legs (**Figure 2B**).

However, a noticeable mean dose reduction is observed in the remaining tissues, bones, and skin, with mean values of 10%, 15%, and 3%, respectively. On the other hand, doses to the penis and testes rise 5% and 8%, respectively. This is due to their proximity to the legs, which results in photons originating in the legs delivering a significant dose to those structures.

The validation of the *leg reconstruction* method for the two FDG-patients confirmed the appropriateness of the operation. In this case, observed differences between incomplete versus complete geometry/source images reach values of ~8% for the RT, 30% for the bones, and 8% for the skin. These differences were reduced to ~-2.5%, ~8%, and ~6% for RT, bones, and skin, when the legs were artificially recreated (**Figure SM2**). The mean doses for testes and penis, which suffered a reduction of ~5% and ~3% in the absence of the legs, were almost fully recovered with the *leg reconstruction* method (differences ≤0.5%).

Three patients, whose whole forearms and hands were not imaged, were simulated with the *arm reconstruction* procedure. The resulting doses (**Figure 2C**) showed a mean reduction in the mean dose to the RT of 5% with respect to the case where the forearms were missing. Bones and skin mean doses also decreased by 11% and 5%, respectively. A slight increase of ~2% in the brain dose is noticed.

It must be pointed out that the relative variability observed in small-size low-uptake tissues, such as ears or lenses, is mainly due to statistical MC uncertainties.

### 3.2. Patient cohort dosimetry: MC calculations versus MIRD

Doses reported in this section correspond to calculations with no bladder emptying after the study and ignore the contribution of missing activities in the legs and arms. Data dispersion among patients is represented with boxplots, where central ticks show the median, boxes range from 25 to 75 percentiles, and whiskers extend to the most extreme data points, excluding outliers (beyond 2.7σ).

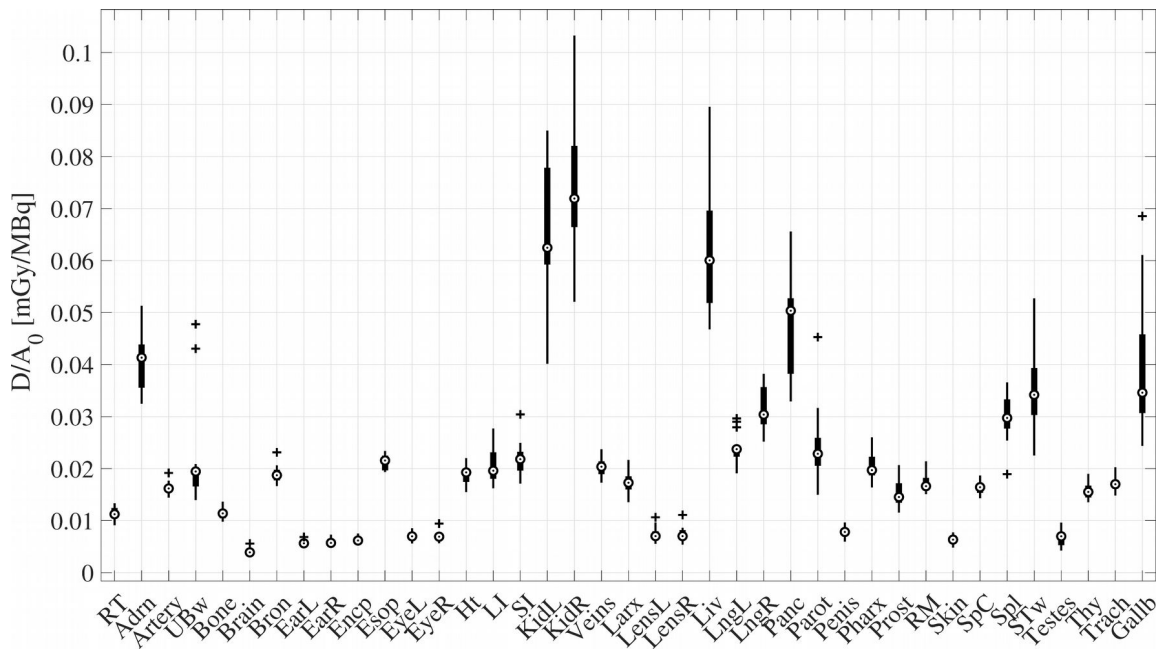

**Figure 3:** Dose per injected activity in our FCH-PET patient cohort. The urinary bladder was not emptied. Boxes range from 25 to 75 percentiles, and whiskers extend to the most extreme data points, excluding outliers (beyond 2.7σ), which are represented by "+" symbols.

We report the dose per injected activity for our patient cohort in **Figure 3.** The highest doses are observed in left and right kidneys, liver, and pancreas, with median values of $6.25\times10^{-2}$ and $7.19\times10^{-2}$, $6.00\times10^{-2}$ and $5.04\times10^{-2}$ mGy/MBq respectively. The gallbladder, adrenals, or stomach wall also present relatively high doses, with medians of $3.46\times10^{-2}$, $4.13\times10^{-2}$, and $3.42\times10^{-2}$ mGy/MBq. The relative uncertainties were lower than 1% for most organs, except for some low-uptake small tissues, such as the lenses and the ears, for which uncertainties were around 1-4 %.

The mean effective dose was calculated from tissue weighting factors recommended by ICRP 103[37]. It reaches $(2.24\pm0.28)\times10^{-2}$ mSv/MBq, which is consistent with that reported by the ICRP 128[6] for the adult male: $2.0\times10^{-2}$ mSv/MBq.

The comparison of mean doses calculated with GATE and the two phantoms shows important discrepancies (**Figure 4**). Relative differences with respect to GATE roughly range in [-37%, +17%] for the CE phantom, and [-33%, +10%] for the ICRP phantom. The most important disagreement is observed for the kidneys in the CE phantom, with a median difference of 33% and a wide range of interpatient variability ([-20%, 64%]).

Discrepancies diminish in some instances when applying mass correction (**Figure 5**). The improvement is especially remarkable for those organs that present high uptake. For instance, the kidneys, where mass correction reduces the differences between MC and phantom-based doses to a median value of -5.6% (range [-12%, +2%]). However, large relative differences are still present for some tissues even after mass scaling.

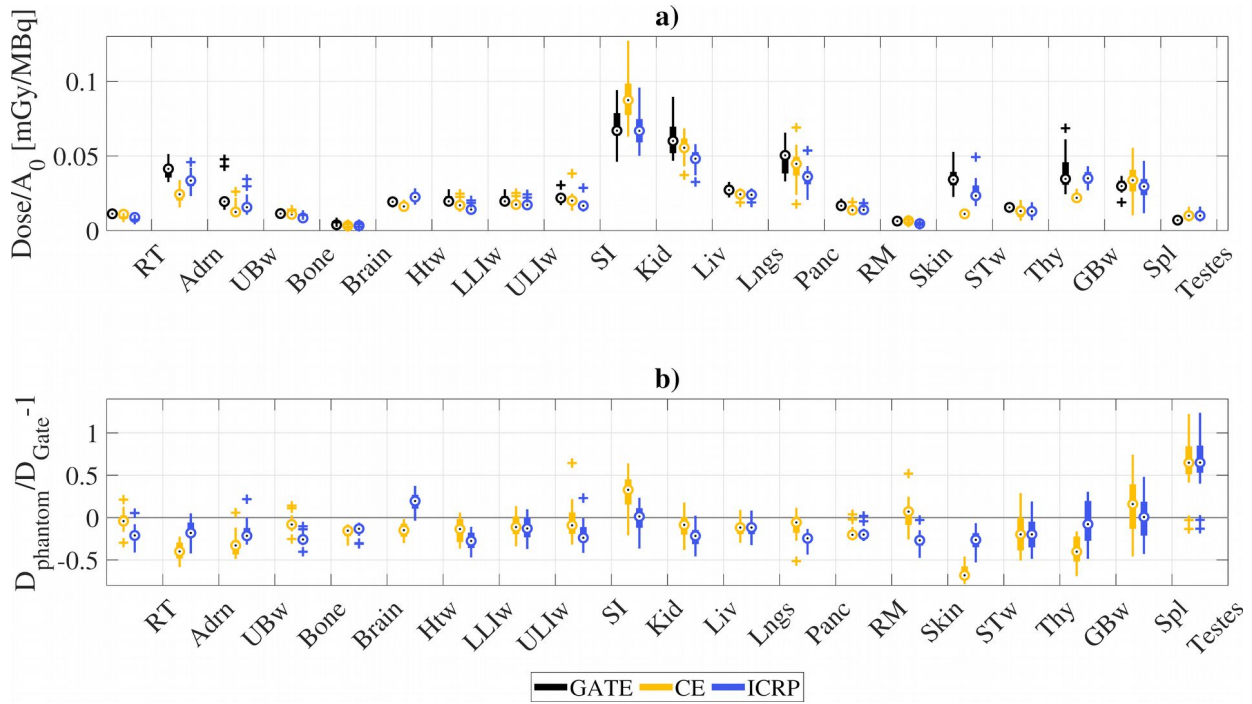

**Figure 4:** Comparison between personalized Monte Carlo and MIRD (Cristy-Eckerman and ICRP's phantom) mean doses per organ. a) Mean dose per injected activity predicted by GATE, CE phantom, and ICRP phantom without the SAFs mass correction. b) Differences between calculations in GATE and both phantoms. Both LLI and ULI are compared against the mean dose of the manually-contoured LI. For the sake of comparison of the ICRP phantom, we considered that the ULI is equivalent to the right colon, and the LLI is approximately equal to the left colon plus the rectum and sigmoid colon. The ICRP phantom doses from the Alimentary and Respiratory systems (including intestines, stomach, and lungs) are reported in the stem cells layers and alveolar interstitium, respectively. Boxes range from 25 to 75 percentiles, and whiskers extend to the most extreme data points, excluding outliers (beyond 2.7σ), which are represented by "+" symbols.

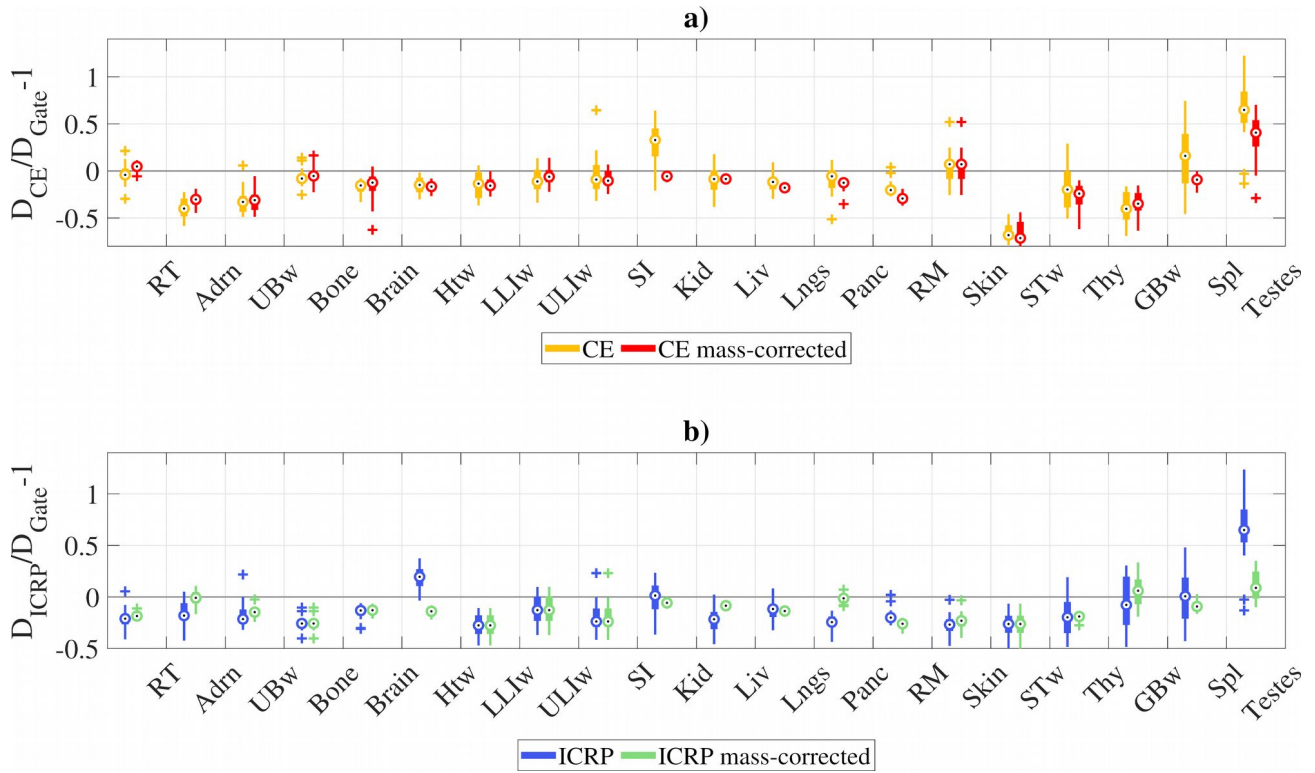

**Figure 5:** Differences between the calculations in GATE and the computational phantoms with and without mass-scaling for a) Cristy-Eckerman phantom; b) ICRP phantom. Boxes range from 25 to 75 percentiles, and whiskers extend to the most extreme data points, excluding outliers (beyond 2.7σ), which are represented by "+" symbols.

Finally, the ICRP 128 dose-per-injected-activity coefficients show important disagreement with the GATE calculations for our patient cohort, as shown in **Figure 6**. Differences for kidney doses are particularly noticeable, with ICRP values of 0.1 mGy/MBq in comparison with a median value of 0.06 mGy/MBq according to GATE, and the urinary bladder with values of 0.06 mGy/MBq (ICRP) versus a median of 0.02 mGy/MBq (GATE). A comparison in terms of residence times, defined as the total number of disintegrations per injected activity, between our cohort and those recommended by the ICRP 128[6] showed good agreement, **Table 1**. An exception was noticed for the urinary bladder contents, which presented remarkably lower cumulated activities.

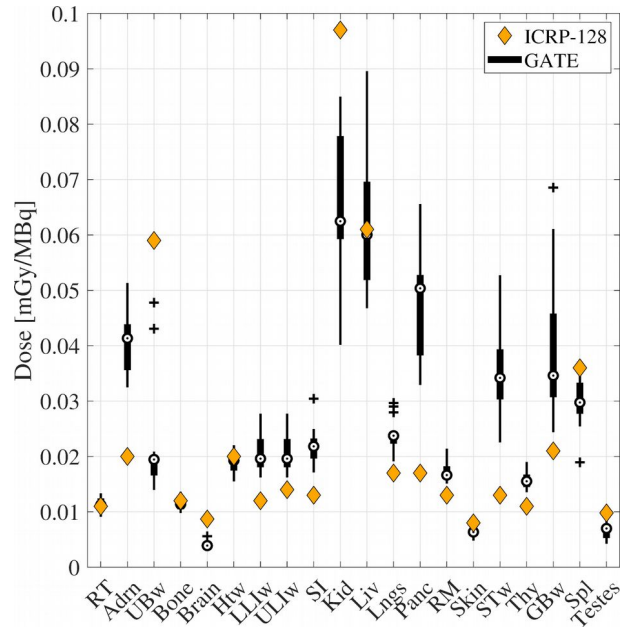

**Figure 6:** Comparison between Monte Carlo computed dose coefficients in our patient cohort and the recommended values provided by the ICRP 128. Boxes range from 25 to 75 percentiles, and whiskers extend to the most extreme data points, excluding outliers (beyond 2.7σ), which are represented by "+" symbols.

| Organs | Residence time, GATE Mean [min, max] (min) | Residence time, ICRP-128 (min) |
|---|---|---|
| Liver | 24.91 [16.04, 31.21] | 25.2 |
| Spleen | 1.52 [0.10, 2.82] | 1.32 |
| Kidneys | 7.95 [5.38, 11.7] | 8.4 |
| Urinary Bladder (w/o elim.) | 0.88 [0.32, 2.44] | 6 |
| Urinary Bladder (w/ elim.) | 0.40 [0.13, 1.26] | 6 |
| Other organs | 100.28 [85.92, 111.65] | 97.8 |

**Table 1.** Residence times of the patient cohort and comparison with those recommended for the FCH by the ICRP-128[6].

# 4. Discussion

We have computed individualized MC doses in a cohort of 17 patients undergoing [18]FCH PET studies, following a methodology that we previously used for FDG[17]. The process includes individualized activity maps and geometries, which are integrated into a MC simulation within the GATE framework. We have compared MC doses to values obtained with widely used phantom-based methods, aiming at examining the adequacy of such non-individualized methods.

Our results show high variability of the relative doses in the patient cohort, with the kidneys and liver being the organs receiving the highest doses, with median values around $\sim 6\times 10^{-2}$ mGy/MBq. Other structures like the stomach wall, pancreas, adrenals, and gallbladder also present moderate to high doses, with median values around $3.5\times 10^{-2}$ mGy/MBq. This is probably partly due to the photonic contribution of the liver to the nearest organs. Non-negligible differences were observed between the right/left lung and kidney. In both cases, the right organ shows higher doses than the left one, which can be caused by a photonic contribution from the liver. The excess dose in the right lung might as well be slightly affected by a diaphragmatic respiratory artifact and partial volume effect (PVE) from the liver. Breathing artifacts are partially avoided in the CT due to the protocol: rapid helix mode acquisition and training of the patient to hold his/her breath, without reaching maximum inspiration capacity, until the end of the CT. However, they may appear in the PET images due to the PET's long acquisition time.

For some organs/tissues, MC doses present discrepancies with the MIRD method if no mass correction is applied. In most cases, the differences were in the range of [-30, 15]%. The mass correction of the S-factors leads to better agreement between MC and MIRD, especially in the case of large organs with high uptake. However, mass correction factors do not improve the agreement for organs with an important photonic contribution to the dose, like the brain, the thyroid, adrenals, or testes. In addition, depending on the phantom, doses to some tissues could not be easily mass corrected. For instance, the human alimentary tract (HAT) of the ICRP phantom reports the S-factors only for the stem cells of the walls. In this case, the use of the mass-scaling relations as described by Stabin et al[11] might be not appropriate. Other cases were also problematic, in particular the hollow organs (with wall and content components). These organs can be difficult to segment as a consequence of the PET/CT resolution limits. In addition, the variation of the S-factor with the mass of the contents/wall, although studied throughout the literature[38,39], does not obey the same relations as those determined for self- and cross-irradiation for most source-target pairs (**Equations (3) and (4)**). As a result, discrepancies may arise in organs such as the intestines or gallbladder.

Finally, the comparison with the dose coefficients recommended in the ICRP 128 showed the impact of patient variability if low levels of individualization are performed in a practical case, resulting in underestimation or overestimation of the dose in several tissues such as the pancreas or kidneys.

We have investigated several sources of uncertainty in computed dose values. Among them, the reconstruction of time-activity-curves from a single (or limited number) of PET scans can be quite important. The effect of this source of uncertainty can be reduced with dynamic PET studies, but it is not clinically feasible to perform dynamic PETs for every single patient to improve the accuracy of internal dosimetry calculations. Nonetheless, the fast FCH uptake by body tissues and slow biological clearance, which has been remarked in several studies, facilitate the biokinetic modeling with a simple exponential fit. The error in terms of the cumulated activities obtained with this method is estimated to be less than ±10% for those organs actively involved in the biokinetics. Uncertainties in absorbed doses associated with the chosen biokinetic model approach should be similar to those estimated for the cumulated activity, especially for organs with high uptake, as shown in our previous study[17].

However, these uncertainties could be higher in the case of the UB, as the mismatch between ICRP-128 and the cohort residence times suggested. The lack of information regarding the emptying scheme complicates an accurate evaluation. An analysis of our cohort of patients showed that the mean fill rate until the end of the PET study was ~0.018% of injected activity per minute, which is compatible with the data collected by Giussani et al[25]. Taking into account that the patients were asked to void their bladder before the PET/CT, an important amount of activity in the bladder was already eliminated by the time the PET/CT was performed, which will introduce a bias in our calculation of bladder doses and residence times. As the residence time recommended by the ICRP corresponds to an emptying scheme of 3.5 h, we consider that this value can be overestimated with respect to our cohort, although without ruling out the possibility of underestimation of our calculations due to our model simplifications.

The comparison between the limit cases for UB emptying shows little effect in doses to most organs. The dose in the UB wall is the most affected, but only for two patients did the differences exceed $1.0 \times 10^{-2}$ mGy/MBq, equivalent to ~2.5 mGy for the typical injected activity. The prostate, which is the second most affected organ, presents changes as large as 0.6 mGy.

We have shown, even if for a limited number of patients, that the use of partial PET/CT images mostly affects the mean doses calculated for remaining tissues, skin, and bones. This effect can be explained by the large contribution of the limbs to those structures, and the low dose the limbs receive. Therefore, if the limbs are not included in the calculation, it leads to an over-estimation of doses to such structures. This limitation could be taken into account by including a dose scaling factor when partial images are acquired, as is often the case.

The application of individualized MC approaches in the clinic still does not look feasible. The main drawbacks are the long calculation times demanding computational resources, the skill- and time-demanding requirements of the segmentation task (and other image processing), and the need to apply clinical protocols that allow to collect more information for biokinetic modeling. Developments to improve these limitations will be paramount to extend the use of individualized dose calculation tools for radiopharmaceuticals.

## 5. Conclusions

A previously described MC method for individualized dosimetry was applied to calculate doses in patients undergoing [18F]FCH PET/CT. The MC-based methodology showed great discrepancies with the MIRD method, independently of the employed phantom. Mass correction of S-factors with patient-specific information allowed for a substantial reduction in the differences in large high-uptake tissues. However, considerable limitations were still present for organs with a higher contribution of photonic dose from cross-irradiation. The disagreement between the MC results and the general recommendations for dosimetry in diagnostic procedures of the ICRP 128 remarked the importance of individualized calculations.

**Ethics approval and consent to participate:** This study was approved by the CEIm-G ethics committee. Informed consent was obtained from all individual participants included in the study.

**Acknowledgments:** This project has received funding from the Instituto de Salud Carlos III (CPII17/00028, and PI17/01428 grants, FEDER co-funding). This project has received funding from the European Union's Horizon 2020 research and innovation programme under the Marie Skłodowska-Curie grant agreement No 839135. We are thankful to Centro de Supercomputación de Galicia (CESGA) for their support, and to Dr. Teresa Cabaleiro and Dr. Ana Estany for their assistance on obtaining ethical clearance for this work.

## References

1. Bauman G, Belhocine T, Kovacs M, Ward A, Beheshti M, Rachinsky I. 18F-fluorocholine for prostate cancer imaging: a systematic review of the literature. *Prostate Cancer Prostatic Dis*. 2012;15(1):45-55. doi:10.1038/pcan.2011.35

2. Langsteger W, Heinisch M, Fogelman I. The role of fluorodeoxyglucose, 18F-dihydroxyphenylalanine, 18F-choline, and 18F-fluoride in bone imaging with emphasis on prostate and breast. *Semin Nucl Med*. 2006;36(1):73-92. doi:10.1053/j.semnuclmed.2005.09.002

3. Hara T, Kosaka N, Kishi H. Development of 18F-fluoroethylcholine for cancer imaging with PET: synthesis, biochemistry, and prostate cancer imaging. *J Nucl Med*. 2002;43(2):187-199.

4. Lam WW-C, Ng DC-E, Wong WY, Ong SC, Yu SW-K, See SJ. Promising role of [18F] fluorocholine PET/CT vs [18F] fluorodeoxyglucose PET/CT in primary brain tumors—Early experience. *Clin Neurol Neurosurg*. 2011;113(2):156-161. doi:10.1016/j.clineuro.2010.09.012

5. European Commission. Council Directive 2013/59/Euratom of 5 December 2013 laying down basic safety standards for protection against the dangers arising from exposure to ionising radiation, and repealing Directives 89/618/Euratom, 90/641/Euratom, 96/29/Euratom, 97/43/Euratom and 2003/122/Euratom. Published online 2014. Accessed September 24, 2020. http://data.europa.eu/eli/dir/2013/59/oj/eng

6. ICRP. ICRP publication 128: Radiation dose to patients from radiopharmaceuticals: a compendium of current information related to frequently used substances. *Ann ICRP*. 2015;44(2S). doi:10.1177/0146645318773622

7. Loevinger R, Budinger TF, Watson EE. *MIRD Primer for Absorbed Dose Calculations*. Society of Nuclear Medicine; 1988.

8. Bolch WE, Eckerman KF, Sgouros G, Thomas SR. MIRD Pamphlet No. 21: A Generalized Schema for Radiopharmaceutical Dosimetry--Standardization of Nomenclature. *J Nucl Med*. 2009;50(3):477-484. doi:10.2967/jnumed.108.056036

9. Xu XG, Bolch WE, Broggio D, et al. The Consortium of Computational Human Phantoms. Published 2005. Accessed September 23, 2020. https://www.virtualphantoms.org/phantoms.htm

10. Snyder WS, Ford MR, Warner GG, Watson SB. MIRD pamphlet no. 11: S absorbed dose per unit cumulated activity for selected radionuclides and organs. *Soc Nucl Med*. Published online 1975.

11. Stabin MG, Sparks RB, Crowe E. OLINDA/EXM: The Second-Generation Personal Computer Software for Internal Dose Assessment in Nuclear Medicine. *J Nucl Med*. 2005;46(6):1023-1027.

12. Petoussi-Henss N, Bolch WE, Zankl M, Sgouros G, Wessels B. Patient-specific scaling of reference S-values for cross-organ radionuclide S-values: what is appropriate? *Radiat Prot Dosimetry*. 2007;127(1-4):192-196. doi:10.1093/rpd/ncm270

13. Marine PM, Stabin MG, Fernald MJ, Brill AB. Changes in Radiation Dose with Variations in Human Anatomy: Larger and Smaller Normal-Stature Adults. *J Nucl Med*. 2010;51(5):806-811. doi:10.2967/jnumed.109.073007

14. Clark LD, Stabin MG, Fernald MJ, Brill AB. Changes in Radiation Dose with Variations in Human Anatomy: Moderately and Severely Obese Adults. *J Nucl Med*. 2010;51(6):929-932. doi:10.2967/jnumed.109.073015

15. Divoli A, Chiavassa S, Ferrer L, Barbet J, Flux GD, Bardies M. Effect of Patient Morphology on Dosimetric Calculations for Internal Irradiation as Assessed by Comparisons of Monte Carlo Versus Conventional Methodologies. *J Nucl Med*. 2009;50(2):316-323. doi:10.2967/jnumed.108.056705

16. Marcatili S, Villoing D, Mauxion T, McParland BJ, Bardiès M. Model-based versus specific dosimetry in diagnostic context: Comparison of three dosimetric approaches. *Med Phys*. 2015;42(3):1288-1296. doi:10.1118/1.4907957

17. Neira S, Guiu Souto J, Díaz Botana P, et al. Quantification of internal dosimetry in PET patients: individualized Monte Carlo vs generic phantom-based calculations. *Med Phys*. 2020;47(9):4574-4588. doi:10.1002/mp.14344

18. O'Donoghue JA. Implications of Nonuniform Tumor Doses for Radioimmunotherapy. *J Nucl Med*. 1999;40(8):1337-1341.

19. Besemer AE, Yang YM, Grudzinski JJ, Hall LT, Bednarz BP. Development and validation of

RAPID: a patient-specific Monte Carlo three-dimensional internal dosimetry platform. *Cancer Biother Radiopharm*. 2018;33(4):155-165. doi:10.1089/cbr.2018.2451

20. Kost SD, Dewaraja YK, Abramson RG, Stabin MG. VIDA: a voxel-based dosimetry method for targeted radionuclide therapy using Geant4. *Cancer Biother Radiopharm*. 2015;30(1):16-26. doi:10.1089/cbr.2014.1713

21. Petitguillaume A, Bernardini M, Broggio D, de Labriolle Vaylet C, Franck D, Desbrée A. OEDIPE, a software for personalized Monte Carlo dosimetry and treatment planning optimization in nuclear medicine: absorbed dose and biologically effective dose considerations. *Radioprotection*. 2014;49(4):275-281. doi:10.1051/radiopro/2014021

22. Marcatili S, Pettinato C, Daniels S, et al. Development and validation of RAYDOSE: a Geant4-based application for molecular radiotherapy. *Phys Med Biol*. 2013;58(8):2491-2508. doi:10.1088/0031-9155/58/8/2491

23. Fedorov A, Beichel R, Kalpathy-Cramer J, et al. 3D Slicer as an image computing platform for the Quantitative Imaging Network. *Magn Reson Imaging*. 2012;30(9):1323-1341. doi:10.1016/j.mri.2012.05.001

24. Chan TF, Vese LA. Active contours without edges. *IEEE Trans Image Process*. 2001;10(2):266-277. doi:10.1109/83.902291

25. Giussani A, Janzen T, Uusijarvi-Lizana H, et al. A Compartmental Model for Biokinetics and Dosimetry of $^{18}$F-Choline in Prostate Cancer Patients. *J Nucl Med*. 2012;53(6):985-993. doi:10.2967/jnumed.111.099408

26. Tavola F, Janzen T, Giussani A, et al. Nonlinear compartmental model of 18F-choline. *Nucl Med Biol*. 2012;39(2):261-268.

27. DeGrado TR, Reiman RE, Price DT, Wang S, Coleman RE. Pharmacokinetics and Radiation Dosimetry of 18F-Fluorocholine. *J Nucl Med*. 2002;43(1):92-96.

28. Fabbri C, Galassi R, Moretti A, et al. Radiation dosimetry of 18F-flurocholine PET/CT studies in prostate cancer patients. *Phys Med.* 2014;30(3):346-351. doi:10.1016/j.ejmp.2013.10.007

29. Janzen T, Tavola F, Giussani A, et al. Compartmental model of $^{18}$F-choline. In: *Medical Imaging 2010: Biomedical Applications in Molecular, Structural, and Functional Imaging*. Vol

7626. International Society for Optics and Photonics; 2010:76261E. doi:10.1117/12.844219

30. Sarrut D, Bardiès M, Boussion N, et al. A review of the use and potential of the GATE Monte Carlo simulation code for radiation therapy and dosimetry applications. *Med Phys*. 2014;41(6Part1):064301. doi:10.1118/1.4871617

31. Carrier J-F, Archambault L, Beaulieu L, Roy R. Validation of GEANT4, an object-oriented Monte Carlo toolkit, for simulations in medical physics. *Med Phys*. 2004;31(3):484-492. doi:10.1118/1.1644532

32. Villoing D, Marcatili S, Garcia M-P, Bardiès M. Internal dosimetry with the Monte Carlo code GATE: validation using the ICRP/ICRU female reference computational model. *Phys Med Biol*. 2017;62(5):1885-1904. doi:10.1088/1361-6560/62/5/1885

33. Chetty IJ, Rosu M, Kessler ML, et al. Reporting and analyzing statistical uncertainties in Monte Carlo–based treatment planning. *Int J Radiat Oncol Biol Phys*. 2006;65(4):1249-1259. doi:10.1016/j.ijrobp.2006.03.039

34. Cristy M, Eckerman KF. *Specific Absorbed Fractions of Energy at Various Ages From Internal Photon Sources. I-VII*. Oak Ridge, TN: Oak Ridge National Laboratory.; 1987.

35. ICRP. ICRP Publication 110: Adult reference computational phantoms. *Ann ICRP*. 2009;39(2). doi:10.1016/j.icrp.2009.09.001

36. ICRP. ICRP Publication 133: The ICRP computational framework for internal dose assessment for reference adults: specific absorbed fractions. *Ann ICRP*. 2016;45(2). doi:10.1177/0146645316661077

37. ICRP. ICRP Publication 103: The 2007 Recommendations of the International Commission on Radiological Protection. *Ann ICRP*. 2007;37(2-4). doi:10.1016/j.icrp.2007.10.003

38. Snyder WS, Ford MR. Estimation of doses to the urinary bladder and to the gonads. In: *Radiopharmaceutical Dosimetry Symposium*. Department of Health, Education and Welfare, Bureau of Radiological Health; 1976:313-349.

39. Andersson M, Minarik D, Johansson L. Improved estimates of the radiation absorbed dose to the urinary bladder wall. *Phys Med Biol*. 2014;59(9):2173-2182. doi:10.1088/0031-9155/59/9/2173

# Supplementary materials

**INTRODUCTION**

This document presents additional information, such as figures, tables, extended methods, and results. The material given here is not essential to understand the developed study but aims to provide a deeper insight into the problem. The references which appear in this document are listed in the main article.

**MATERIALS AND METHODS**

**Legs reconstruction**

From the last slice of the thighs, new tissues (labeled as bone, skin and RT) were defined. The new RT was created by extruding and thinning the leg tissue (simulating the thickness variation from thigh to ankle), creating a cone-like shape. The bone structure of that last slice was repeated in each newly generated slice, while the most external layer between tissue and air was automatically assigned to skin tissue. The artificial legs were extended until the total height of the patient matched the reported value in the PET/CT metadata. The new HU and activity (without UB emptying) were assigned according to the mean values of the skin and remaining tissue of the last four original leg slices. In **Fig. SM1** we report a slice-cut of the rebuilt legs in the three types of images.

**Arms reconstruction**

For the majority of the patients in the cohort, the PET/CT images did not include the whole arms. In the most extreme case, the forearms were missing for the patient, although generally they could be partially imaged. To characterize the effect in dose of the missing activity we run a set of simulations where the forearms for three patients were roughly approximated. We re-wrote the geometry-source images with a soft-tissue (RT) semi-torus which connected both arm sections and surrounded by a layer of skin. A second concentric semi-torus was defined as bone tissue. The rebuilt forearms were conformed by approximately 3.3 kg of RT, 800 g of bones, and 300 g of skin. The activity and HU associated with the new structure were taken from the mean values in the last three original PET/CT slices. The results of the simulation allowed us to quantify the uncertainties which might arise as a consequence of the absence of part of the arms tissue. However, it has to be taken into account that the chosen reconstruction represented a limit case, where activity and masses are probably slightly over-estimated.

**Exponential versus biokinetic model**

Experimental data of the evolution of activities in several organs with time reported in Giussani et al. **[25]** were extracted with the software tool Engauge Digitizer (markummitchell.github.io/engauge-digitizer). Such data was used to estimate differences between cumulated activities obtained with Giussani's model and a biexponential model (both of them using multiple time-activity points), and a mono-exponential fit, considering only one time-activity point, as in our study. In the latter case, the second activity-point (t≈75 min) reported in Giussani's work was chosen as the reference value.

Finally, we paid attention to the blood contribution to the total CA. It was already demonstrated that the injected activity quickly disappears from the blood. Integration of the blood activity concentration from Degrado et al. (Table 2) **[27]** yielded reports of a relative activity concentration Ã≈10% in the blood in the first 30 minutes after the injection. Data from Giussani et al **[25]**, who warned of a possible overestimation of the blood compartment, established a relative contribution of ~10 %. However, in their study non-realistic blood volumes were obtained for their cohort (as high as ~25 liters). If we integrate the blood-TACs in Giussani's Supplemental Materials for patients with early blood examples (patients 5, 6, 8, 9, and 10), taking a more plausible blood volume of 5 liters, it can be seen that the blood contribution could be as low as ~3.5%. This value seems to support that activity in the blood will not affect much internal dose distributions.

**Supplementary Tables**

**Table SM1**. List of abbreviations of the segmented organs

| Abbreviation | Organ/Tissue | Abbreviation | Organ | Abbreviation | Organ |
|---|---|---|---|---|---|
| Adrn | Adrenals | KidR | Right Kidney | RT[b] | Remaining Tissues |
| Artery | Large Arteries | Larx | Larynx | SI | Small Intestine |
| Bone | Bones | LensL | Left lens | Skin | Skin |
| Brain | Brain | LensR | Right lens | SpC | Spinal Cord |
| Bron | Bronchial tree | LI | Large Intestine[a] | Spl | Spleen |
| Brst | Breasts | Liv | Liver | STc | Stomach |

| | | | | | contents |
|---|---|---|---|---|---|
| EarL | Left Ear | LngL | Left lung | STw | Stomach wall |
| EarR | Right Ear | LngR | Right lung | Testis | Testis |
| Encp | Encephalic trunk | Ovary | Ovaries | Thy | Thyroid |
| Esop | Esophagus | Panc | Pancreas | Trach | Trachea |
| EyeL | Left Eye | Parot | Parotids | UBc | Urinary Bladder contents |
| EyeR | Right Eye | Penis | Penis | UBw | Urinary Bladder wall |
| Gall | Gallbladder | Pharx | Pharynx | Uter | Uterus |
| Ht | Whole Heart | Prost | Prostate | Veins | Large Veins |
| KidL | Left Kidney | RM | Red Marrow | | |

[a] The Large Intestine was manually segmented as one structure in our cohort. When compared with Cristy-Eckerman phantom, it is considered to be composed of the sum of Low Large Intestine (LLI) and Upper Large Intestine (ULI). For the ICRP phantom, it is considered to be formed by the Right Colon (RC), Left Colon (LC) and, Recto-Sigmod (RS).

[b] Remaining Tissues are those organs or tissues which belong to the Whole Body but were not explicitly segmented.

**Supplementary Figures**

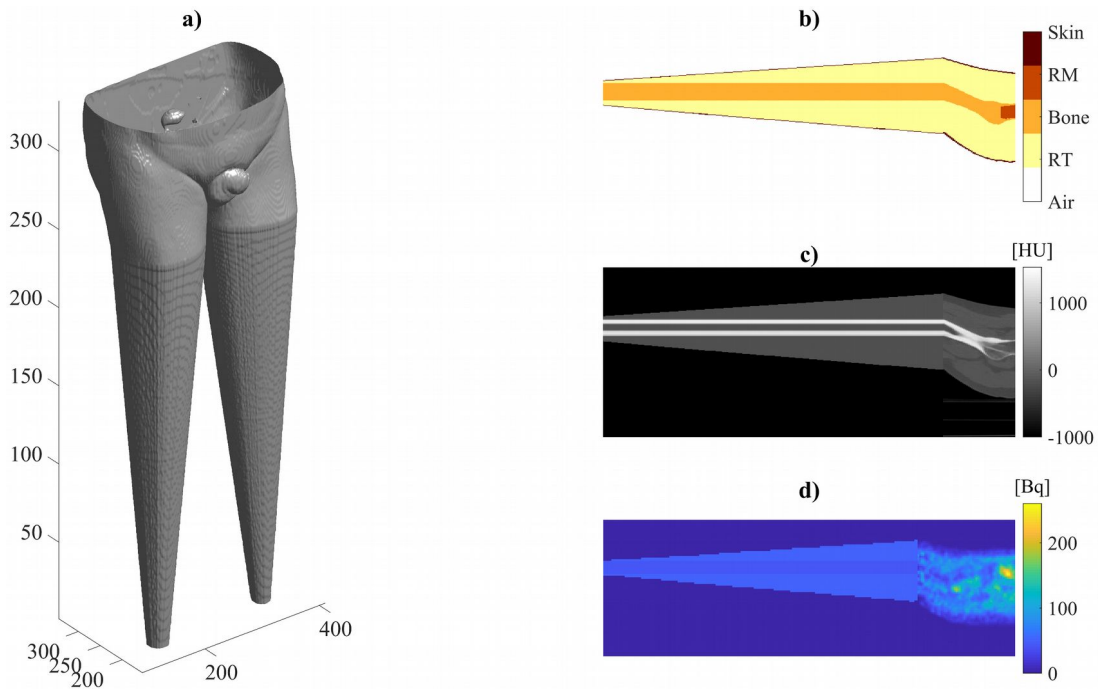

Figure SM1: Example of reconstructed legs for a 3D representation of geometry (a) and one slice of (b) the segmentation map, (c) Hounsfield units map, and (d) activity map.

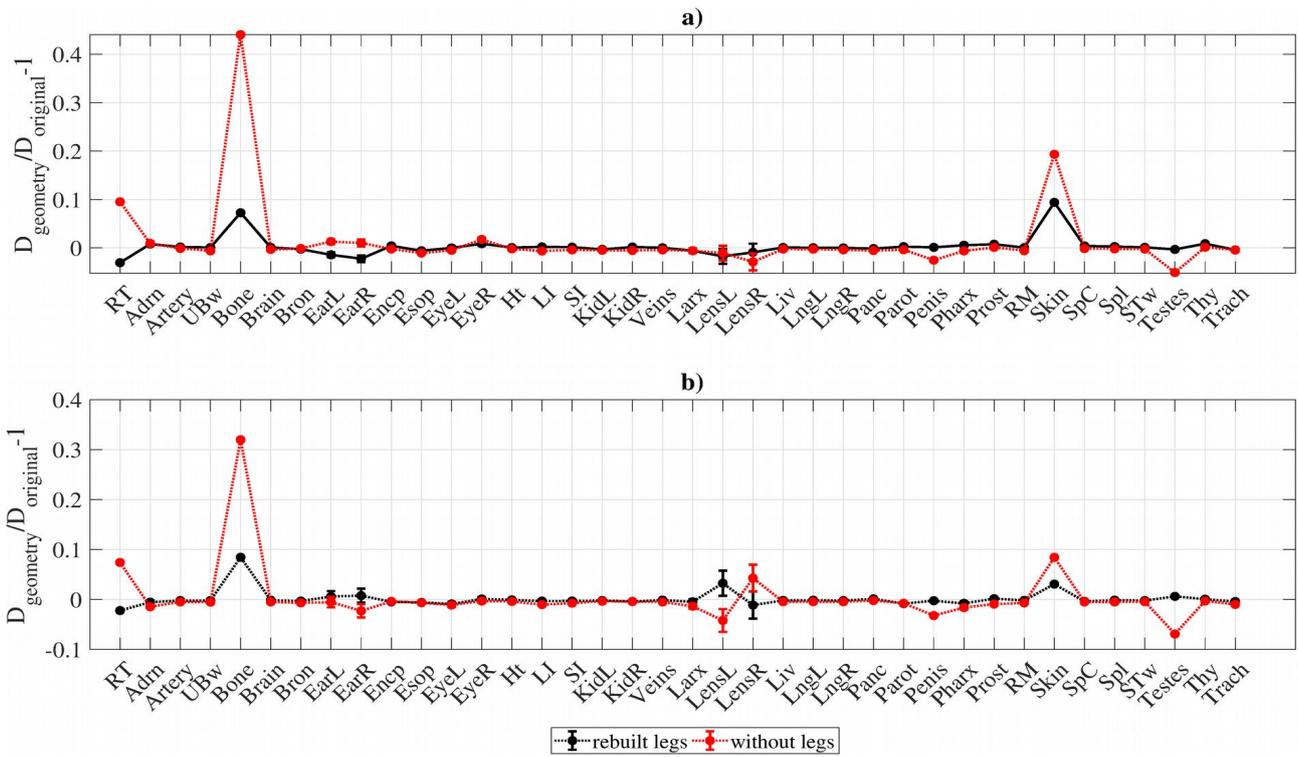

Figure SM2: Validation of the *leg reconstruction* with two (a and b) whole-body FDG-PET/CT patients taken from [17]. Doses calculated in the patient when removing the legs differ up to 40% for bone and 20% for skin. Such differences are importantly reduced when adding the *leg reconstruction* methodology (<10%).